\documentclass[
twocolumn,
superscriptaddress,
amsmath,amssymb,
aps,
prd,
]{revtex4-2}

\usepackage{ctex}
\usepackage{lipsum}
\usepackage{array}
\usepackage{graphicx}
\usepackage{xcolor}
\usepackage{amsmath}
\usepackage{booktabs}
\usepackage{multirow}
\usepackage{natbib}
\setcitestyle{numbers,comma,square}
\usepackage{lineno}
\usepackage{comment}

\usepackage{subfigure}
\usepackage[colorlinks=true, urlcolor=black, citecolor=blue, linkcolor=blue]{hyperref}

\begin{document}

\title{GPS-Synchronized Monitoring of Core-collapse Supernova Bursts with PandaX-4T via Coherent Elastic Neutrino Nuclear Scattering
}

\def\tdli{State Key Laboratory of Dark Matter Physics, Key Laboratory for Particle Astrophysics and Cosmology (MoE), Shanghai Key Laboratory for Particle Physics and Cosmology, Tsung-Dao Lee Institute \& School of Physics and Astronomy, Shanghai Jiao Tong University, Shanghai 201210, China}
\def\sjtuphys{State Key Laboratory of Dark Matter Physics, Key Laboratory for Particle Astrophysics and Cosmology (MoE), Shanghai Key Laboratory for Particle Physics and Cosmology, School of Physics and Astronomy, Shanghai Jiao Tong University, Shanghai 200240, China}
\def\newcorner{New Cornerstone Science Laboratory, Tsung-Dao Lee Institute, Shanghai Jiao Tong University, Shanghai 201210, China}
\def\MESJTU{School of Mechanical Engineering, Shanghai Jiao Tong University, Shanghai 200240, China}
\def\SPEIT{SJTU Paris Elite Institute of Technology, Shanghai Jiao Tong University, Shanghai 200240, China}
\def\SJTUSC{Shanghai Jiao Tong University Sichuan Research Institute, Chengdu 610213, China}

\def\BUAA{School of Physics, Beihang University, Beijing 102206, China}
\def\BUAACenter{Peng Huanwu Collaborative Center for Research and Education, Beihang University, Beijing 100191, China}
\def\BUAALab{International Research Center for Nuclei and Particles in the Cosmos \& Beijing Key Laboratory of Advanced Nuclear Materials and Physics, Beihang University, Beijing 100191, China}
\def\SCNT{Southern Center for Nuclear-Science Theory (SCNT), Institute of Modern Physics, Chinese Academy of Sciences, Huizhou 516000, China}

\def\USTClab{State Key Laboratory of Particle Detection and Electronics, University of Science and Technology of China, Hefei 230026, China}
\def\USTCdep{Department of Modern Physics, University of Science and Technology of China, Hefei 230026, China}

\def\YaLongSD{Yalong River Hydropower Development Company, Ltd., 288 Shuanglin Road, Chengdu 610051, China}
\def\scKeyLab{Jinping Deep Underground Frontier Science and Dark Matter Key Laboratory of Sichuan Province, Liangshan 615000, China}

\def\pku{School of Physics, Peking University, Beijing 100871, China}
\def\CHEPpku{Center for High Energy Physics, Peking University, Beijing 100871, China}

\def\SDUdep{Research Center for Particle Science and Technology, Institute of Frontier and Interdisciplinary Science, Shandong University, Qingdao 266237, China}
\def\SDUlab{Key Laboratory of Particle Physics and Particle Irradiation of Ministry of Education, Shandong University, Qingdao 266237, China}

\def\UMD{Department of Physics, University of Maryland, College Park, Maryland 20742, USA}

\def\SYU{School of Physics, Sun Yat-Sen University, Guangzhou 510275, China}
\def\SYUSFI{Sino-French Institute of Nuclear Engineering and Technology, Sun Yat-Sen University, Zhuhai 519082, China}
\def\SYUzhuhai{School of Physics and Astronomy, Sun Yat-Sen University, Zhuhai 519082, China}
\def\SYUshenzhen{School of Science, Sun Yat-Sen University, Shenzhen 518107, China}

\def\NKU{School of Physics, Nankai University, Tianjin 300071, China}
\def\YTU{Department of Physics, Yantai University, Yantai 264005, China}
\def\FDU{Key Laboratory of Nuclear Physics and Ion-beam Application (MOE), Institute of Modern Physics, Fudan University, Shanghai 200433, China}
\def\CDUT{College of Nuclear Technology and Automation Engineering, Chengdu University of Technology, Chengdu 610059, China}

\affiliation{\tdli}
\author{Binyu Pang}\affiliation{\SDUdep}\affiliation{\SDUlab}
\author{Zihao Bo}\affiliation{\sjtuphys}
\author{Wei Chen}\affiliation{\sjtuphys}
\author{Xun Chen}\affiliation{\tdli}\affiliation{\SJTUSC}\affiliation{\scKeyLab}
\author{Yunhua Chen}\affiliation{\YaLongSD}\affiliation{\scKeyLab}
\author{Chen Cheng}\affiliation{\BUAA}
\author{Xiangyi Cui}\affiliation{\tdli}
\author{Manna Deng}\affiliation{\SYUSFI}
\author{Yingjie Fan}\affiliation{\YTU}
\author{Deqing Fang}\affiliation{\FDU}
% ...
\collaboration{PandaX Collaboration}
\noaffiliation

\maketitle

\vskip 1.5mm

\onecolumngrid

The landmark detection of neutrinos from SN1987A marked the dawn of neutrino astrophysics. The neutrino burst provided essential insights into fundamental properties of neutrinos, and served as key probes of stellar evolution and supernova dynamics. The recent advancement in coherent elastic neutrino-nucleus scattering enables the detection of core-collapse supernova burst neutrinos using tonne-scale liquid xenon detectors originally designed for dark matter direct detection. Leveraging this capability, we developed and deployed an online supernova monitoring system for the PandaX-4T experiment. This system features a GPS module with millisecond-level timing precision, a low false-alarm rate, and high sensitivity to galactic core-collapse supernova explosion events. The methodology is robust, directly scalable, and planned for implementation in the next-generation PandaX-20T experiment.
\twocolumngrid
\vskip 4mm

The core-collapse supernova (CCSN) explosions occur fewer than three times per century within the Milky Way~\cite{1}. The CCSN bursts occur in the latter stages of evolution of massive stars with masses exceeding eight times that of the Sun. Roughly $10^{53}$ erg of energy is released during a CCSN burst, with neutrinos carrying away about $99\%$ of the energy~\cite{8}. The neutrino burst lasts only a few or a few tens of seconds, and unlike electromagnetic radiation or gravitational waves, neutrinos can escape the stellar core more easily and are emitted earlier, allowing them to reach ground-based observatories before other messengers~\cite{9}. The only supernova (SN) neutrinos ever detected were from SN1987A, with around 20 events observed by a few detectors~\cite{2,3,4}. These neutrino events provided critical insights into the fundamental properties of neutrinos, and inspired theoretical models on CCSN explosion mechanisms~\cite{add4,add5,add6}. However, progress in the field remains hindered by the lack of statistics of SN neutrinos. To prepare for neutrino detection from the next galactic CCSN burst, several neutrino experiments have implemented specialized supernova monitoring systems~\cite{5,6,7,add7}, forming a global organization known as the Supernova Early Warning System (SNEWS)~\cite{10,11}.

Recently, significant advancements have been made in neutrino detection methods. In 2017, with neutrinos from a spallation neutron source, the COHERENT experiment reported the first observation of coherent elastic neutrino-nucleus scattering (CE$\nu$NS)~\cite{12}, a coherently enhanced weak interaction predicted by Freedman in 1974~\cite{Freedman}. CE$\nu$NS offers a remarkable opportunity for detecting low energy astrophysical neutrinos~\cite{CEvNSNB,CEvNnobel,CEvNSXe,add13,add14}, as its cross-section scales approximately with the square of the target nucleus's neutron number, $\sigma_{\nu}\simeq N^{2}G^{2}_{\text{F}}E^{2}_{\nu}/4\pi$, where $N$ is the neutron number, $G_\text{F}$ is the Fermi coupling constant, and $E_\nu$ is the neutrino energy~\cite{13}. Furthermore, as a neutral current process, CE$\nu$NS is sensitive to all neutrino flavors, providing a valuable complement to experiments that primarily focus on specific flavors~\cite{17,18}. The challenge, however, lies in the extremely low nuclear recoil energy, which is difficult to detect. Over the past two decades, liquid xenon-based dark matter direct detection experiments, designed to detect rare and low-energy scattering events, have made remarkable progress in improving their sensitivities. Experiments such as XENONnT, LUX-ZEPLIN (LZ), and XMASS~\cite{6,14,xmass} have proposed leveraging CE$\nu$NS to detect SN neutrinos. Notably, in 2024, the PandaX-4T~\cite{add13} and XENONnT~\cite{add14} collaborations independently reported indications of CE$\nu$NS from solar $^8$B neutrinos with xenon nuclei, and very recently, the LZ collaboration also reported the evidence of $^8$B CE$\nu$NS with more than 4-$\sigma$ significance~\cite{add14p}. These findings mark a significant milestone towards the realization of SN neutrino detection via CE$\nu$NS. 
In this paper, we will discuss the performance of a newly developed CCSN burst monitoring system in PandaX-4T, 
utilizing online selection of CE$\nu$NS events.

The PandaX-4T experiment is located in the China Jinping Underground Laboratory (CJPL), shielded by 2,400 meters of rock overburden. Comprehensive details about the detector can be found in Ref.~\cite{19}, with only a few essential aspects highlighted in this Letter. The primary detector is a cylindrical xenon time projection chamber (TPC) containing approximately 3.7 tonnes of liquid xenon (LXe). A physical event produces a delayed coincidence of photons from prompt scintillation (S1) and delayed electroluminescence photons from ionized electrons (S2), which are detected via the top and bottom photomultiplier (PMT) arrays. The waveforms from all fired PMTs are read out under a continuous and $\lq\lq$triggerless" data acquisition mode, and event building and reconstructions are performed by analysis software. The three-dimension interaction vertex is reconstructed using the top PMT pattern and the time difference between S1 and S2, allowing for further suppression of external background through fiducialization.

The PandaX-4T has been in operation since November 2020 and is scheduled to continue the data taking until the end of 2025. To allow an online monitoring for potential CCSN explosions, a dedicated GPS-based CCSN burst system was established in July 2025. About 83 days of physics data collected between July 18 and October 14, 2025, will be used to evaluate its performance.
 
\begin{figure*}[t] 
\centering
  \includegraphics[scale=0.35]{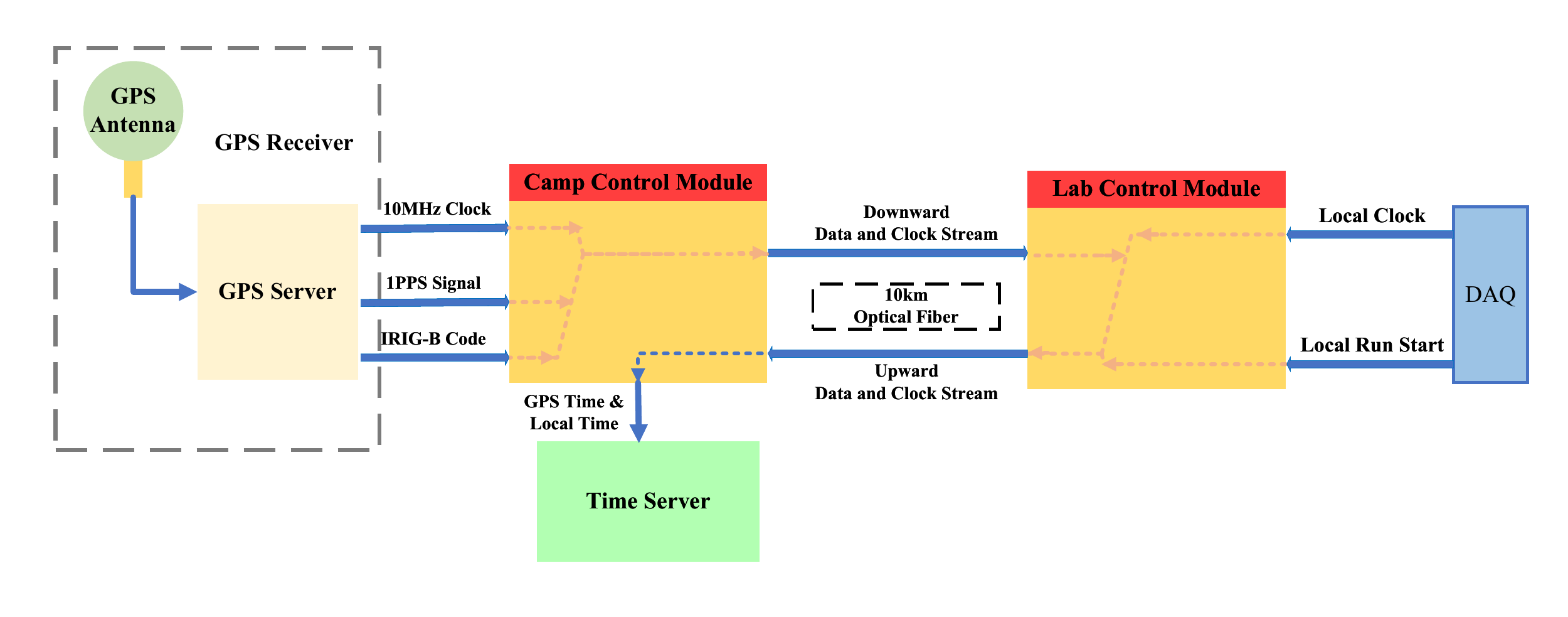}
\caption{\label{GPStime} Schematic diagram of the PandaX-4T's GPS-based timing system, which comprises a GPS receiver, a camp control module, a lab control module and a time server. The mapping of the delivered GPS time and the DAQ hardware clock ticks is stored in a database. See text for more details. } 
\end{figure*}

A  millisecond-level GPS-based timing system was specially designed for PandaX-4T, as shown in Fig.~\ref{GPStime}, to facilitate collaborative early warnings with other neutrino experiments via time-delay triangulation~\cite{11,triangulation}. The GPS ground sub-system comprises a GPS receiver and a camp control module. The GPS antenna receives satellite signals and transmits them to the GPS server, which generates a high-precision 10 MHz reference clock through its internal clocking mechanism to synchronize the local clock of the camp control module. The precise GPS time, encoded in the IRIG-B code and synchronized with the one pulse per second (PPS) signal, is transmitted to the camp control module and then downwards the control module in the underground laboratory (hereafter referred to as lab control module) through a multi-kilometer optical fiber network. A hardware clock counter of the lab control module synchronizes with the DAQ's master clock at the start of each new data acquisition run (labeled as local clock and local run start in Fig.~\ref{GPStime} ), and the mapping between the delivered GPS time and the hardware clock ticks is stored in a database of a time server. The time delay introduced by the fiber optic length is calculated by measuring the time difference between the emission and reception of synchronized signals sent in opposite directions (downward and upward time streams), using the camp and lab control modules. This delay is utilized to correct timestamps during data acquisition. The precision of this system is maintained at the millisecond level, which is determined primarily by the GPS server.

\begin{figure}[h] 
	\centering
  \centering
  \includegraphics[scale=0.4]{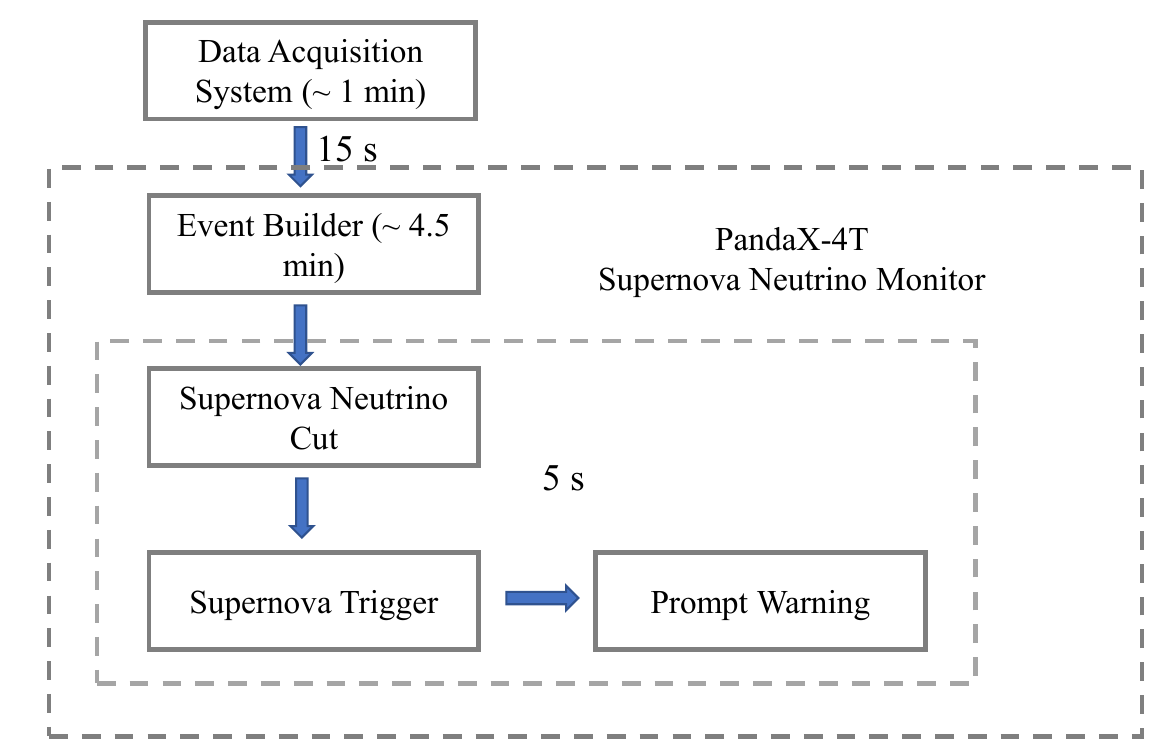}
  \caption{\label{fig:SNdiagram}Schematic diagram of SN online monitoring system in PandaX-4T. The time consumed in each step is shown. }
\end{figure}

We developed an online software-based monitoring system to identify CCSN bursts by detecting transient increases in CE$\nu$NS-like event rate in PandaX-4T induced by SN neutrinos. The flow diagram is shown in Fig.~\ref{fig:SNdiagram}. When the data acquisition system generates a new raw data file, the acquired data is stored using the Bamboo-shoot3 library~\cite{add19,add20}. Each raw file contains approximately one minute of data, with a size of about 2 GB. The data compression process reduces it to about 1 GB and writes the data to local disk, which takes about ten seconds.

\begin{figure*}[t]
	\centering
	
	\subfigure{
		\includegraphics[width=0.65\linewidth]{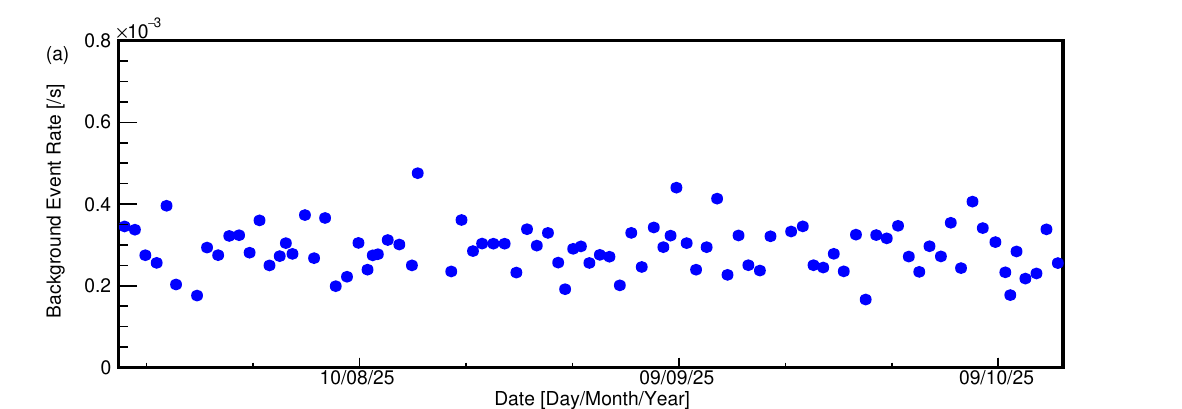} 
		\label{rate_vs_time_a}
	}
	\subfigure{
		\includegraphics[width=0.65\linewidth]{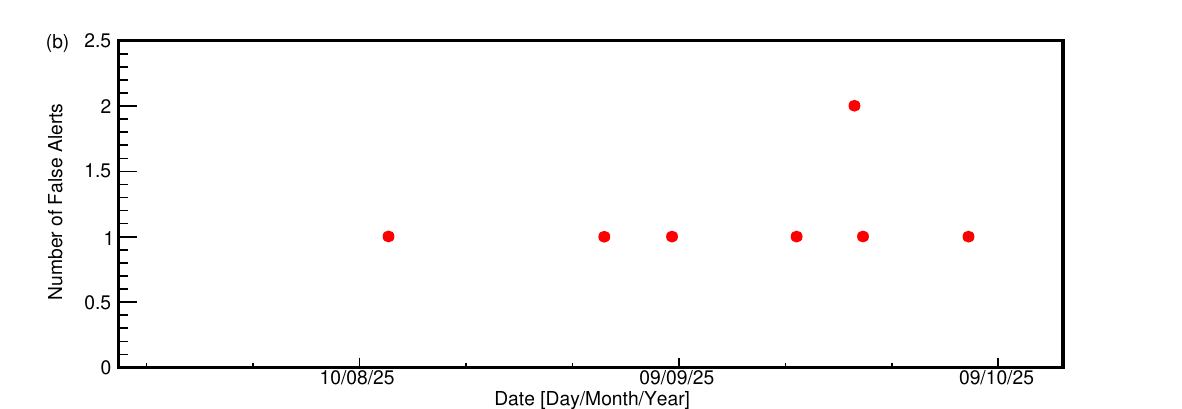}
		\label{rate_vs_time_b}
	}
	\caption{The background event rate (a) and the false alert (b) as a function of time.}
	\label{rate_vs_time}
\end{figure*}

The data is then transmitted to the remote data center in Chengdu, enabling parallel processing of different files. The transmission time is approximately fifteen seconds. Offline processing and event reconstruction take about 4.5 minutes. After reconstruction, event information such as S1 and S2 variables, reconstructed vertex, and other relevant parameters is extracted. A series of cuts are then applied to reduce background, including the afterglow cut (which removes periods immediately following large signals amounting to a 11\% of deadtime), the fiducial volume (FV) cut, the accidental coincidence (AC) cut, and quality cuts for S1 and S2. 
Data quality selection mentioned above is inherited from the previous dark matter analysis~\cite{25}, with minor adjustments implemented to enhance detection sensitivity for supernova neutrino analysis. To select CE$\nu$NS events, the region-of-interest (ROI) for S1 and S2 are set to [2.1, 100] PE and [80, 4000] PE, respectively, to balance the false alert rate due to background contributions and detection efficiency of CE$\nu$NS. The remaining events are considered “single” SN neutrino candidates. Over our 83-day dataset, the single-event rate is on average $2.9 \times 10^{-4} $ $\text{s}^{-1}$, as shown in Fig.~\ref{rate_vs_time_a}. The temporal variation of the rate is correlated with the change of operation conditions of the radon distillation tower~\cite{distillation}.

\vskip 2mm

Finally, a CCSN burst alert is generated if the number of neutrino candidates exceeds the threshold $N_{\text{thr}}$ ($N_{\text{thr}}=2$) within a 10-second time window. Further details on this method are provided in Ref.~\cite{22}. The SN alert information, including the alert start time and the total number of neutrino candidates, is sent to the PandaX-4T SN group via email. This process takes approximately five seconds. Considering all latency factors discussed above, the entire monitoring process takes about 6 minutes, as also indicated in Fig.~\ref{fig:SNdiagram}.

The operational status of the SN monitoring system is displayed on a webpage, allowing personnel to inspect key performance parameters. The system operates stably during routine data-taking periods, with exceptions occurring only during special circumstances, such as power outages or server maintenance. SN monitoring program also automatically excludes calibration data based on the data type defined by the data acquisition system.

To evaluate the efficiency of the single event selection described above, we calculate the expected neutrino energy spectrum using the so-called Garching model with progenitor masses of $M_\text{p} = 11.2M_{\odot}$ and $M_\text{p} = 27M_{\odot}$ ($M_{\odot}$ is the solar mass), employing the LS220 nuclear equation of state (EoS)~\cite{20,21}. The electroweak neutral current process allows most neutrinos to interact coherently with nucleons in the nucleus for neutrino energies $E_{\nu} \lesssim 100$ MeV. Details of the differential cross-section calculation can be found in Ref.~\cite{22}.

The signal response model for PandaX-4T is constructed using the Noble Element Simulation Technique (NEST) v2.3.6 parameterization~\cite{23}, with relevant parameters tuned by fitting calibration data~\cite{add26}. The NEST model simulates the light yield ($L_{\text{y}}$, number of photons per keV) and charge yield ($Q_{\text{y}}$, number of electrons per keV) for nuclear recoil events in LXe. The S1 and S2 signals are then generated using a waveform simulation (WS) package, which assembles real waveforms from the data with $L_{\text{y}}$ and $Q_{\text{y}}$ as inputs~\cite{24}. These WS events are reconstructed and selected using the same procedure as the real experimental data, allowing the overall detection efficiency to be determined.

The overall efficiency includes contributions from signal reconstruction, data quality selection, and the ROI cut, evaluated with the methodology outlined in Refs.~\cite{22,darkmatter}. As an example, the overall detection efficiency for SN neutrino events using the Garching model with $M_\text{p} = 27~M_{\odot}$ is approximately 13\%. The most significant efficiency loss arises from the fact that the majority of CE$\nu$NS events have the low S1 and S2 values.

Based on the $N_{\text{thr}}=2$ threshold mentioned above, eight alert candidates were observed during the 83 days of physics data. The time distribution of these alerts is shown in Fig.~\ref{rate_vs_time_b}. Two alarms were identified in a same day, with a time difference of about 3 hours. No other anomalies were identified after a thorough check.

The false alert background, dominated by accidental coincidences within the 10-second time frame, can be estimated from the rate of “single” SN neutrino candidates. The single rate, $2.9 \times 10^{-4}$ $ \text{s}^{-1}$ (see  Table~\ref{tab:bkrate}), is dominated by electron recoil events, which, as in previous analyses~\cite{25}, arise from material and environmental radioactivity, accidental coincidences, and radioactive noble contaminants (e.g., $^{222}$Rn). The expected false alert rate is approximately 2.2 per month, resulting in 6.1 false alerts in the dataset. Further verification of the consistency between the expected and observed alerts was performed by running our alert algorithm over neutron calibration data from deuteron-deuteron (DD) and AmBe sources, with total live time  of about 2.8 and 3.7 days, respectively. The corresponding “single” SN neutrino candidates rate is approximately $2.2 \times 10^{-2}$ ($9.7 \times 10^{-3}$)~$\text{s}^{-1}$. In these datasets, 886 (288) alerts were observed, compared to the expected 870.6 (265.3) alerts, as estimated using the method in Ref.~\cite{22}. Good agreement was observed between the expected and measured results.

Based on this analysis, we conclude that within our 83-day dataset, the alerts are consistent with the absence of any real CCSN bursts above the background at the 1$\sigma$ level. It is important to note that the primary goal of SNEWS is to search for coincidences between multiple experiments, thereby significantly enhancing the sensitivity to CCSN signals beyond what is achievable by a single experiment. Notably, the false alert rate requirement for an experiment to join SNEWS is one per week\rule[0.5ex]{1.em}{0.65pt}a criterion that PandaX-4T already satisfies.

\renewcommand\arraystretch{1.5}
\begin{table}[htbp]
	\setlength{\abovecaptionskip}{0cm}
	\setlength{\belowcaptionskip}{0.2cm}
	\setlength{\tabcolsep}{3pt}
	\centering
	\caption{\label{tab:bkrate} The observed and expected false alerts are compared using datasets from DD, AmBe, and physical data as detailed in fourth and fifth column. The second and third column present the “single”  SN neutrino candidates rate and the calendar time of the dataset, respectively. }

     \begin{tabular}{ccccc}
        \hline
		Data type & Rate [/s] & Calendar time & Observed & Expected \\
		\hline
		DD& 2.2$\times$$10^{-2}$ & 2.8 days & 886 & 870.6 \\
		AmBe& 9.7$\times$$10^{-3}$ & 3.7 days & 288 & 265.3 \\
		Physical & 2.9$\times$$10^{-4}$ & 83 days & 8 & 6.1 \\	
		\hline
	\end{tabular}
\end{table}

For all eight alerts, intended alert messages to SNEWS were also generated, with one example alert shown in Table~\ref{tab:SNtier}. More details about the SNEWS alert system can be found in Ref.~\cite{26}.

For different (in total five) tiers of alerts, identified by the first column of Table~\ref{tab:SNtier}, SNEWS provides different message template for participating experiments. For each tier, the second and third columns provide essential information for that example PandaX-4T alert.


\begin{table*}[t]
	\centering
	\footnotesize
	\caption{List of different message tiers within the SNEWS network, using a PandaX-4T alert as an example. The detector name is required in each tier and is not listed in the column. 
    For the Coincidence tier, the neutrino event time is represented in UTC format, derived from GPS time in PandaX-4T. 
    For the Significance tier, under the background only assumption, the p-values are the probabilities of observing more than or equal to the number of candidate events 
    within a 5-second time window at the single rate of $2.9 \times 10^{-4}$ $ \text{s}^{-1}$. 
    For the Timing tier, the time series is an ordered sequence of nanosecond-precision integers, representing time offsets relative to the first neutrino candidate.  For the Heartbeat tier, the detector status parameter is set to ON/OFF according to its operational condition. 
    For the Retraction tier, the retraction is initiated via a dedicated command provided by the SNEWS network, depending on real-time experimental considerations. 
    }
    
	\label{tab:SNtier} 
	\begin{tabular}{ccc}
		\hline
		Tier & Required information & Value \\
		\hline
		Coincidence & initial neutrino time & 2025-10-06T14:13:20.333181 \\
		Significance & p-values for time bins, width of time bins & [6.11e-06, 1.00], 5.00 	\\
		Timing & initial neutrino time, neutrino time series  & 2025-10-06T14:13:20.333181, [0, 1213145668]  \\
		Heartbeats & detector status  & ON/OFF	\\
		Retraction & retract latest & \rule{0.5cm}{0.5pt} \\
        \hline 
	\end{tabular}
\end{table*}

\vskip 2mm 

Based on the performance of the SN monitoring system described above, the detection potential of various CCSN explosions with PandaX-4T-only data can be evaluated. The differential event rate of CE$\nu$NS can be expressed as
\begin{equation} \label{eq1}
\begin{split}
	\frac{dN^{2}}{dE_{\text{nr}}dt_{\text{pb}}}(E_{\text{nr}})=&\frac{m_{\text{det}}N_{\text{A}}}{M_{\text{A}}(4\pi d^{2})} \int_{E^{\text{min}}_{\nu}}^{\infty}	 \frac{d\sigma}{dE_{\text{nr}}}(E_{\nu},E_{\text{nr}}) \\
   &\times \Phi(E_\nu, t_{\text{pb}})\epsilon (E_{\text{nr}})dE_{\nu},
\end{split}
\end{equation}
where $t_{\text{pb}}$ represents the time after the CCSN core bounce, $m_{\text{det}} = 2.6$ tonnes is the effective target mass in this analysis, $N_\text{A}$ is Avogadro's number, and the molar mass of xenon is $M_\text{A} = 131.2\, \text{g/mol}$, and $d$ is the distance from the CCSN to Earth. $E^{\text{min}}_{\nu} = \sqrt{m_{\text{A}} E_{\text{nr}}/2}$ represents the minimum neutrino incident energy required to induce a nuclear recoil energy of $E_{\text{nr}}$ in CE$\nu$NS. In practice, we integrate over $E_{\nu} \in (E^{\text{min}}_{\nu},300$ MeV). $d\sigma/dE_{\text{nr}}$ and $\Phi(E_\nu, dt_{\text{pb}})$ are the differential cross-section of the CE$\nu$NS process and the total neutrino fluxes of all flavors at the energy of $E_{\nu}$ and time of $t_{pb}$, respectively.  $\epsilon(E_{\text{nr}})$ represents the detection efficiency. 

\begin{figure}[h] 
	\centering
	\includegraphics[scale=0.35]{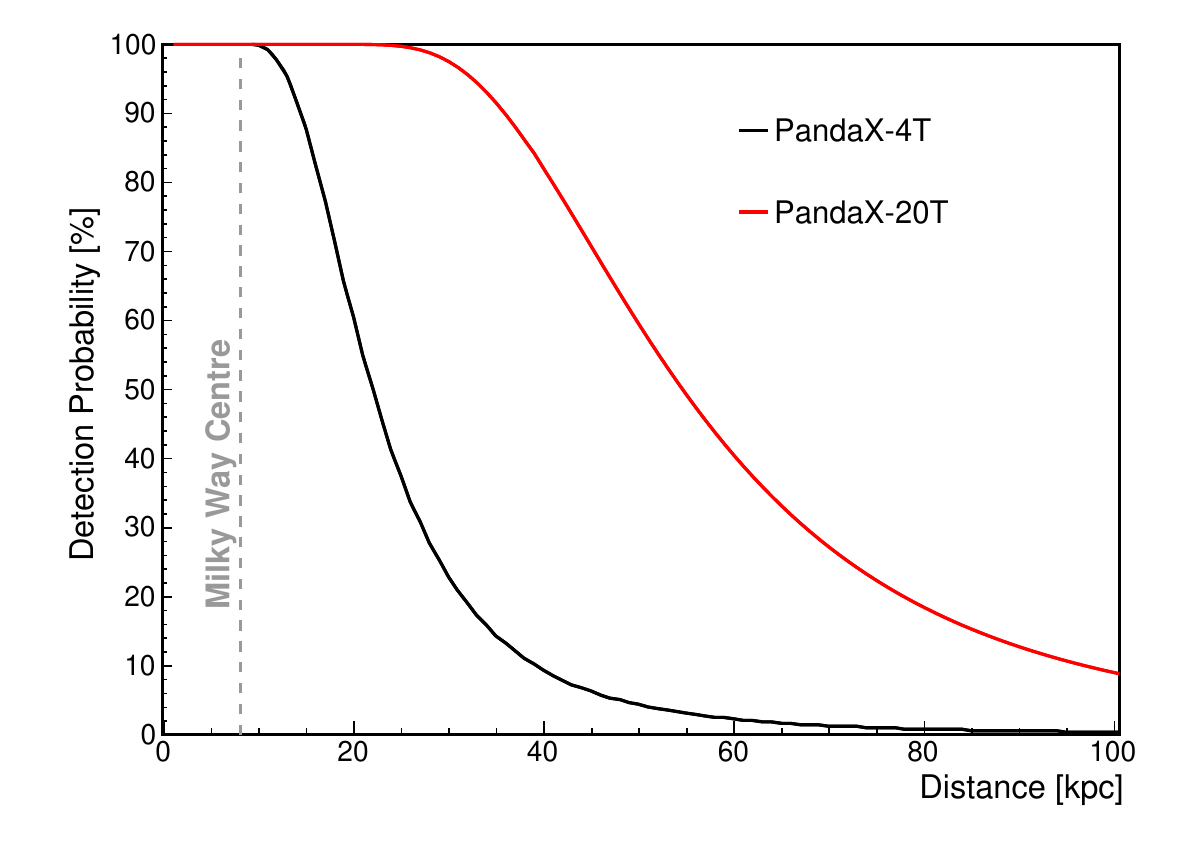}
	\caption{\label{fig:SNprob}Detection probability of the CCSN explosions as a function of distance using the Garching model with $M_{p}=27~M_{\odot}$ for PandaX-4T (black) and PandaX-20T (red). The corresponding false alert rate is twice a month in PandaX-4T.}
\end{figure}

The expected number of SN neutrino events in the detector can be calculated by integrating Eq.~\ref{eq1}. Here, we integrate the variable $t_{\text{pb}}$ for the model with $M_\text{p} = 27~M_{\odot}$ ( and $M_\text{p} = 11.2~M_{\odot}$ ) over the time interval from the onset time of core bounce to 8.35 (7.60) seconds. $E_{\text{nr}}$ is integrated up to 30 keV. The results are summarized in Table~\ref{tab:l2}. The two used typical distances of 10 kpc and 168 pc correspond to the positions of the nearby Galactic center and Betelgeuse~\cite{add33} as observed from Earth, respectively. The detection probability of CCSN explosions, defined as the fraction of the observed over the total, is estimated using the Garching model with $M_{p}=27~M_{\odot}$, employing the LS220 EoS. Figure~\ref{fig:SNprob} illustrates the detection probability as a function of CCSN distance to the PandaX-4T and the next generation experiment PandaX-20T, where we assume that PandaX-20T has a fiducial mass of $\sim 16$ tonnes. PandaX-4T achieves nearly-$100\%$ detection efficiency at 10 kpc, while the PandaX-20T maintains $\sim$100\% efficiency up to 30 kpc.

\renewcommand\arraystretch{1.5}
\begin{table}[htbp]

	\centering
	\caption{\label{tab:l2} The number of expected SN neutrinos for source distances of 10 kpc and 168 pc in PandaX-4T, using Garching models ($M_{p}=11.2~M_{\odot}$ and $M_{p}=27~M_{\odot}$) with the LS220 EoS. The fourth column indicates the time interval from the onset time of core bounce to the model's termination time.}

     \begin{tabular}{cccc}
        \hline	   
		SN model & d=10 kpc & d=168 pc & Duration [s] \\
		\hline
		11.2 $M_{\odot}$ & 3.7 & 1.3$\times 10^{4}$  & 7.60 \\
		27 $M_{\odot}$ & 8.1 & 2.9 $\times 10^{4}$ & 8.35 \\
		\hline
	\end{tabular}
\end{table}

\begin{figure*}[t]
	\centering

	\subfigure{
		\includegraphics[width=0.45\textwidth]{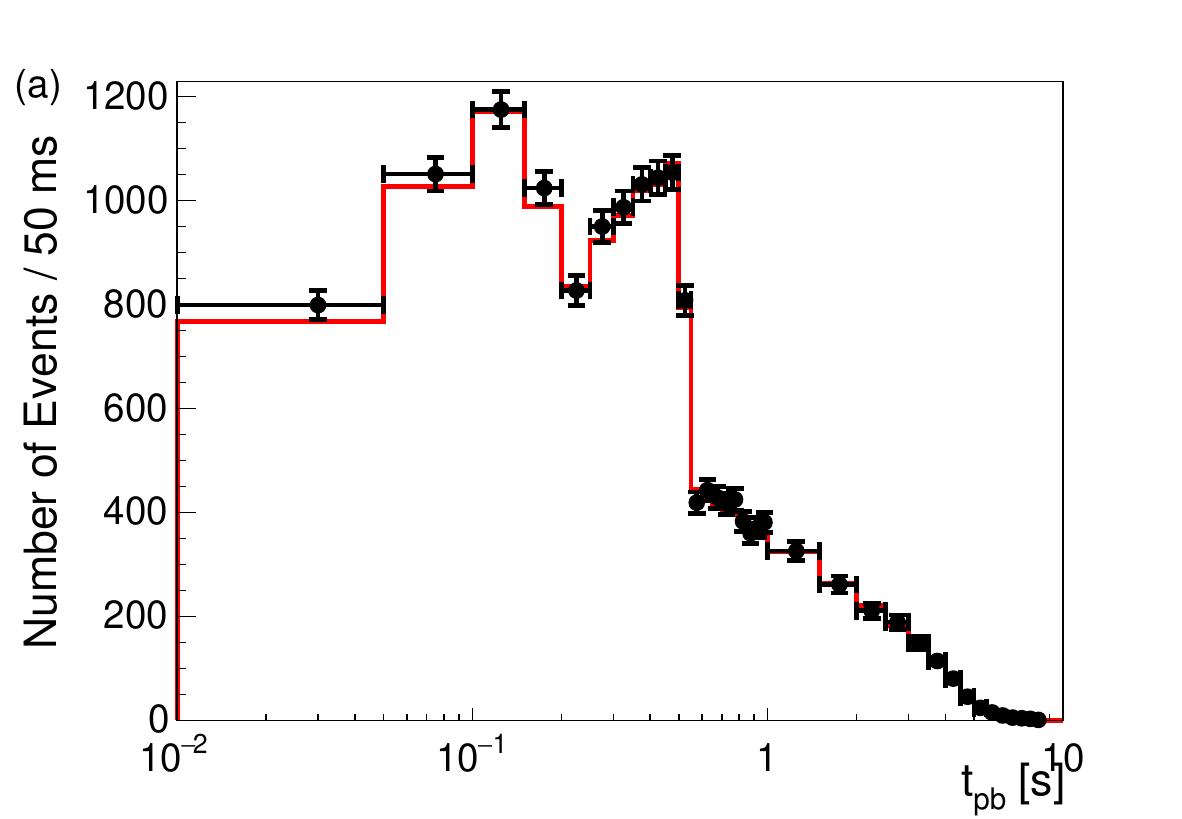}
		\label{snspectrum_a} 
	}
	\subfigure{
		\includegraphics[width=0.45\textwidth]{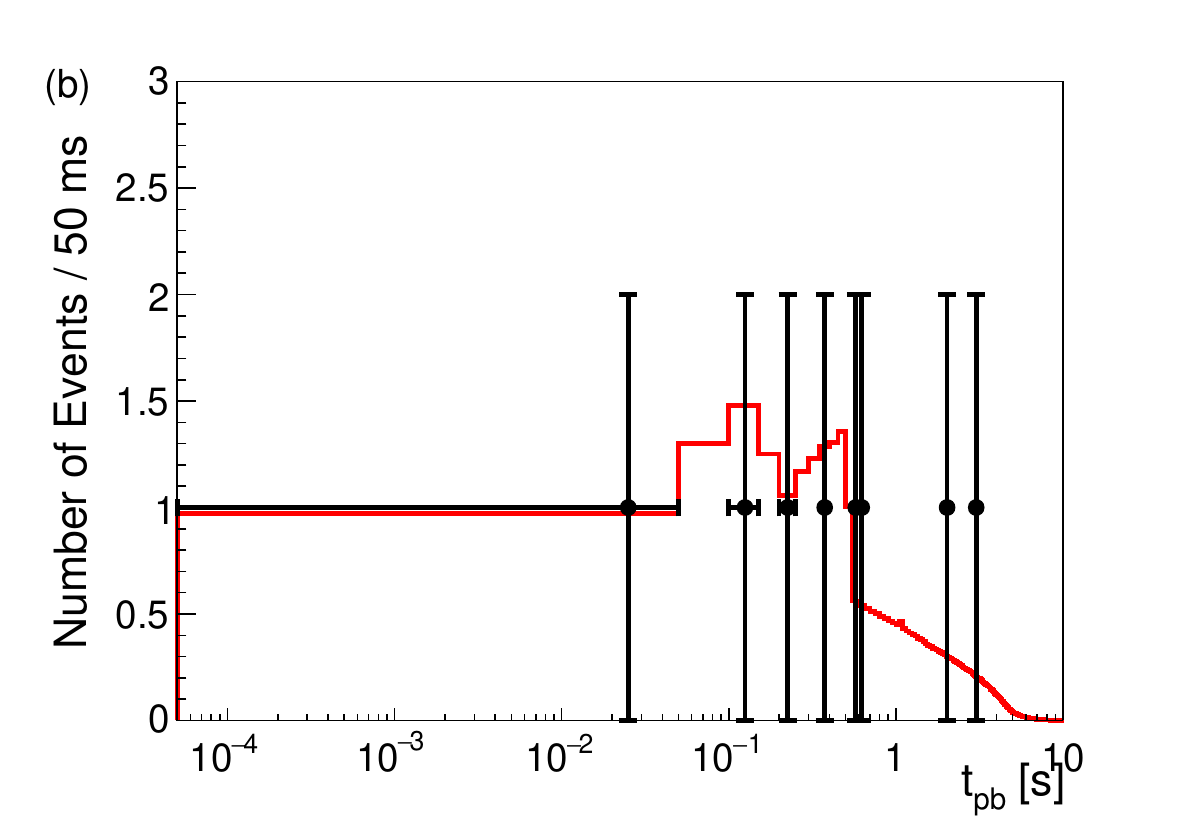}
		\label{snspectrum_b}
	}
	\par\vspace{-0.95\baselineskip}

	\subfigure{
		\includegraphics[width=0.45\textwidth]{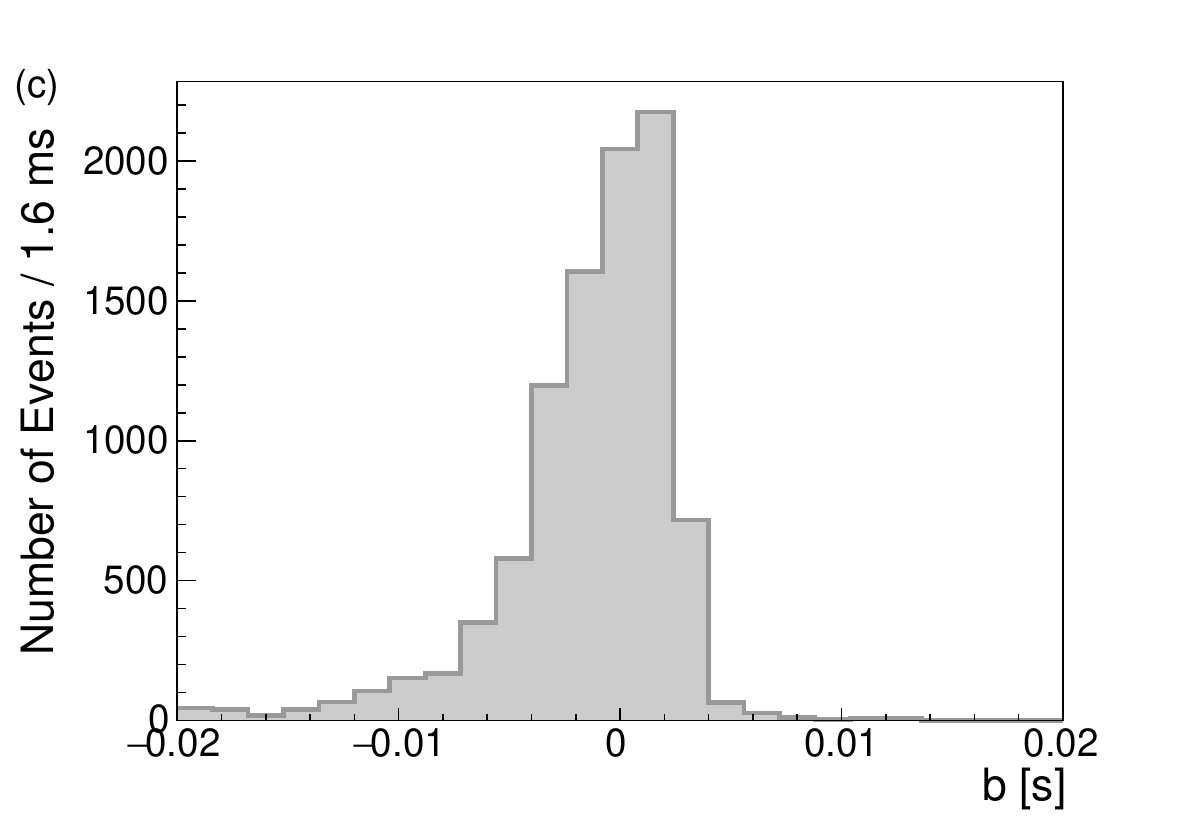}
		\label{snspectrum_c} 
	}
	\subfigure{
		\includegraphics[width=0.45\textwidth]{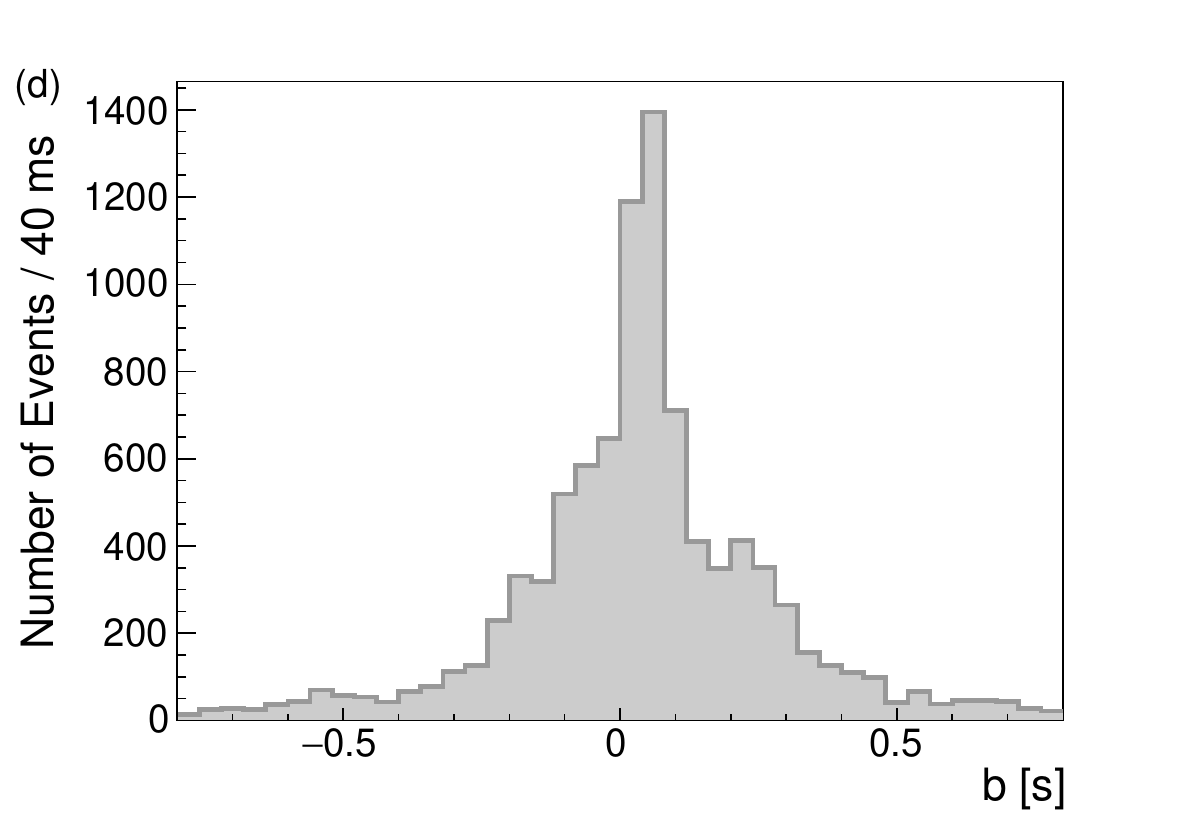}
		\label{snspectrum_d}
	}

		\caption{\label{snspectrum}Typical timing distributions of the detected SN neutrino events from the distance of 168 pc (a) and 10 kpc (b). Data points are produced by simulations and the red lines are the fit results.  The fitted onset times relative to the true time in about 10,000 mock detections are shown in the bottom panels. The resulting onset time measurement exhibits a variation of $[-2.7, 2.4]$ ms (c) and $[-0.38, 0.11]$ s  (d) relative to zero for 168 pc and 10 kpc, respectively, within a 68\% confidence level (C.L.).}
		
\end{figure*}

The core bounce onset time can also be determined when sufficient event statistics are available. A typical fitting result of the timings after the core bounce of the SN neutrino events from the distance of 168 pc is displayed in Fig.~\ref{snspectrum_a}. The points represent the simulated SN neutrino events, and the histogram represents the fitting result using the equation $a \cdot f(t_{\text{pb}}-b)$, where $f$ denotes the time profile function of the CCSN events, while $a$ and $b$ are fit parameters. 
To estimate how well we can determine the onset time, signified by the precision on the parameter $b$, we repeat the fits for many copies of mock time profiles, with results shown in Figs.~\ref{snspectrum_c} and \ref{snspectrum_d}. The 68\% C.L. intervals for the variation of the offset times relative to the true value are $[-2.7, 2.4 ]$ ms at 168 pc and $[-0.38, 0.11]$ s at 10 kpc due to smaller statistics. Note that the described fitting method may not represent the timing precision achievable for promptly sharing data from a real supernova, as the true burst profile may not match the input model. As an alternative, we also consider the core bounce onset time as the arrival time of the first neutrino candidate, based on sampling from the Garching model. The resulting 68\% C.L. intervals for the onset time variation are approximately 1 ms at 168 pc and 0.2 s at 10 kpc. The uncertainty at 168 pc is dominated by GPS timing precision. 


In summary, we have established a robust online supernova monitoring system with GPS time synchronization implemented in PandaX-4T, leveraging the novel detection channel of coherent elastic neutrino-nucleus scattering. The performance of this system during an 83-day data-taking period, from July 18, 2025, to October 14, 2025, is presented. During this period, eight alerts were produced, consistent with a false alert rate of 2.2 per month caused by background. This monitoring system has the capability to provide the astronomical community with early warnings of CCSN explosions with a latency of about 6 minutes, and achieves nearly $100\%$ detection efficiency for supernovae occurring within 10 kpc of Earth in PandaX-4T. The system is ready to join the SNEWS network. 

Looking ahead, the implementation of this system in the next-generation PandaX-20T experiment is expected to significantly enhance sensitivity to core-collapse supernovae within our galaxy. By enabling precise measurements of the recoil energy spectrum and time profile features, it offers an exciting opportunity to significantly advance our understanding of the fundamental properties of neutrinos and provide critical insights for refining astrophysical models.

Acknowledgment: This project is supported in part by grants from National Key R\&D Program of China (Nos. 2023YFA1606200, 2023YFA1606203), National Science Foundation of China (Nos. 12090060, 12090063, U23B2070, 12175139), and by Office of Science and Technology, Shanghai Municipal Government (grant Nos. 21TQ1400218, 22JC1410100, 23JC1410200, ZJ2023-ZD-003). We thank for the support by the Discipline Construction Fund of Shandong University, and the Fundamental Research Funds for the Central Universities. We also thank the sponsorship from the Chinese Academy of Sciences Center for Excellence in Particle Physics (CCEPP), Thomas and Linda Lau Family Foundation, New Cornerstone Science Foundation, Tencent Foundation in China, and Yangyang Development Fund. Finally, we thank the CJPL administration and the Yalong River Hydropower Development Company Ltd. for indispensable logistical support and other help.


\end{document}